\documentclass[preprintnumbers,amsmath,amssymb]{revtex4}
\usepackage[utf8x]{inputenc}
\usepackage{ucs}

\usepackage{amsfonts}
\usepackage{dcolumn}
\usepackage{bm}
\usepackage[usenames]{color}
\usepackage{multirow}
\usepackage{graphicx}
\usepackage{hyperref}
\usepackage{subfig}
\usepackage{booktabs}
\usepackage{xcolor}
\usepackage[normalem]{ulem}

\usepackage{float} 
\usepackage{wrapfig} 

\usepackage{upgreek} 
\usepackage{cancel} 
\usepackage{mathdots} 
\usepackage{mathrsfs} 
\graphicspath{{figures/}} 

\begin{document}

\preprint{draft version}

\title{Entropy and complexity properties of the $d$-dimensional blackbody radiation}

\author{I. V. Toranzo$^{1}$ and
J. S. Dehesa$^{1}$}

\address{
  $^{1}$Instituto Carlos I de F\'{\i}sica Te\'orica y
Computacional and Departamento de F\'{\i}sica At\'omica Molecular
y Nuclear, Universidad de Granada, Granada 18071, Spain.
}

\begin{abstract}
Space dimensionality is a crucial variable in the analysis of the structure and dynamics of natural systems and phenomena. The dimensionality effects of the blackbody radiation has been the subject of considerable research activity in recent years.
These studies are still somewhat fragmentary, posing formidable qualitative and quantitative problems for various scientific and technological areas. In this work we carry out an information-theoretical analysis of the spectral energy density of a d-dimensional blackbody at temperature T by means of various entropy-like quantities (disequilibrium, Shannon entropy, Fisher information) as well as by three (dimensionless) complexity measures (Crámer-Rao, Fisher-Shannon and LMC). All these frequency-functional quantities are calculated and discussed in terms of temperature and dimensionality. It is shown that all three measures of complexity have an universal character in the sense that they depend neither on temperature nor on the Planck and Boltzmann constants, but only on the the space dimensionality d. Moreover, they decrease when d is increasing; in particular, the values 2.28415, 1.90979 and 1.17685 are found for the Crámer-Rao, Fisher-Shannon and LMC measures of complexity of the 3-dimensional blackbody radiation, respectively. In addition, beyond the frequency at which the spectral density is maximum (which follows the well-known Wien displacement law), three further characteristic frequencies are defined in terms of the previous entropy quantities; they are shown to obey Wien-like laws. The potential usefulness of these distinctive features of the blackbody spectrum is physically discussed.
\vskip
0.5cm

\noindent
Keywords: d-dimensional physics, information theory, black-body radiation, cosmic microwave background, Planck distribution, Wien’s law, disequilibrium, Shannon entropy, Fisher information, Crámer-Rao complexity, Fisher-Shannon complexity, LMC complexity, Heisenberg frequency, Shannon frequency, Fisher frequency.\\

PACS: 06.30.Bp, 42.50.Ct, 44.40.+a, 05.70.-a
\end{abstract}

\maketitle

\section{Introduction}

Space dimensionality plays a very fundamental role in numerous scientific and technological areas, from field theory, string theory and quantum cosmology (see e.g., \cite{green,weinberg}), to atomic and molecular physics (see e.g., \cite{herschbach}), quantum optics \cite{rybin}, condensed matter (see e.g., \cite{acharyya,march,xluo}) and quantum information and computation (see e.g.,\cite{krenn, spengler}). See also \cite{dehesa1} for a recent summary. The idea that the universe is trapped on a membrane in some high-dimensional space-time may explain why gravity is so weak, and could be tested at high-energy particle accelerators \cite{cicoli, sandes} in the ongoing LHC experiments with hadronic beams colliding at 7 to 14 TeV. On the other hand, many quantum systems and phenomena possess natural generalizations in which the number of degrees of freedom is a free parameter \cite{dehesa2,harrison, maia,salen}. On the other hand, it is well-known that the dimensionality substantially modifies the physical solutions of the quantum wave equations of the systems, and thus all their properties (see e.g., \cite{witten,dehesa2,dong, brandon}).\\

In the last few years there has been a growing interest in the analysis of the influence of the space dimensionality in the blackbody radiation\cite{cardoso,ramos1,alnes,lehoucq,stewart,nozari,zeng,ramos2} from different standpoints. This is not surprising at all. Let us mention, for instance, that the cosmic microwave background radiation today is known to be the most perfect blackbody radiation ever observed in nature, with a temperature of about 2.725 K \cite{mather0}.\\

Various research \cite{cardoso, ramos1} has shown that the spectral energy density of a d-dimensional ($d > 1$) blackbody at temperature T (i.e., the energy per frequency and volume units contained in the frequency interval ($\nu$, $\nu + d\nu$) inside a d-dimensional enclosure maintained at temperature T) is given  by the generalized Planck radiation law
\begin{equation}
\rho_{T}^{(d)}(\nu) = \frac{2 (d-1) h \left(\frac{\sqrt{\pi }}{c}\right)^d}{\Gamma \left(\frac{d}{2}\right) }\frac{\nu^{d}}{e^{\frac{h \nu }{k_{B} T}}-1}, 
\end{equation}
where T denotes the blackbody temperature, $h$ and $k_B$ are the Planck and Boltzmann constants, respectively, and $\Gamma(x)$ denotes the gamma function of Euler \cite{olver}.
Up until now there is no information-theoretic analysis of this unimodal density in the literature, to the best of our knowledge. This is certainly striking not only from a mathematical standpoint, but mainly because of its physical relevance in so many areas of physics, chemistry and technology as briefly mentioned above. In this study we carry out such an analysis by means of the main entropy quantifiers (disequilibrium, Shannon entropy and Fisher information) and the following three complexity measures of intrinsic type: Crámer-Rao, Fisher-Shannon and Lopez-Ruiz-Mancini-Calvet (LMC in short).\\

The structure of the work is the following. In section II the basic spreading quantities (variance, entropy and complexity measures) of a general continuous one-dimensional probability distribution are defined, and their meanings and properties relevant to this effort are briefly given and discussed. In Section III the variance and the main entropy-based quantifiers (disequilibrium, Shannon entropy and Fisher information) of the d-dimensional blackbody spectrum are explicitly determined in terms of temperature and space dimensionality. Later, in section IV the subsequient results are used to define three new characteristic spectral frequencies besides the frequency $\nu_{max}$ at which the spectrum reaches its maximum, finding that they obey some displacement law similar to the well-known Wien law followed by $\nu_{max}$. Then, in section V the measures of complexity of the blackbody spectrum are examined, finding that they depend on the space dimensionality only; this dependence is numerically studied. Finally, some concluding remarks are given, and various open problems are pointed out.

\section{Information-theoretic measures: Basics}

In this Section we briefly describe the definitions and meanings of the entropy and complexity measures of a probability distribution. Let us consider a general one-dimensional random variable $X$ characterized by the continuous probability distribution $\rho(x)$,
$x \in \Lambda \subseteq \mathbb{R}$. Obviously it is asumed that the density is normalized to unity, so that $\int_{\Lambda} \rho(x) dx  =  1$. To quantify the spread of $X$ over the interval $\Lambda$ we usually employ the statistical root-mean-square
or standard deviation (or even, Heisenberg length)  $l_{Heis} \equiv\Delta x$, which is the square root of the variance
   \begin{equation*}
      V[\rho]=\langle x^2 \rangle-\langle x \rangle^2,
   \end{equation*}
where
\[
\langle f(x) \rangle= \int_{\Lambda} f(x) \rho(x) dx.
\]
   
The information theory provides other spreading measures such as the disequilibrium, the Shannon entropy and the Fisher information. The Shannon entropy $S[\rho]$ of $\rho(x)$ is defined \cite{shannon_49} by
  \begin{eqnarray}
  \label{eq:shann}
     S[\rho] = -\int_{\Lambda} \rho(x) \ln \rho(x) dx.
  \end{eqnarray}
The disequilibrium (also known as the Onicescu entropy) \cite{onicescu} is given by 
\begin{eqnarray}
\label{eq:diseq}
D[\rho] &=&\langle \rho \rangle =\int_{\Lambda}\rho(x)^{2}\, dx 
\end{eqnarray}

The Fisher information of $\rho(x)$ is defined \cite{frieden_04,fisher} as
  \begin{eqnarray}
  \label{eq:fisher}
     F\left[\rho\right] =\int_{\Lambda} \frac{\left(\frac{d}{dx} \rho(x)\right)^2}{\rho(x)}\,dx.
  \end{eqnarray}

It is worth remarking that: (a) these three information-theoretic spreading measures do not depend on any particular point of their
interval $\Lambda$, contrary to the standard deviation, (b) the Fisher information has a locality property because it is a functional 
of the derivative of $\rho(x)$, and (c) the standard deviation and the disequilibrium and Shannon entropies are global properties because
they are power and logarithmic functionals of $\rho(x)$, respectively. Moreover they have different units, so that they can not be compared each other. To overcome this difficulty, the following information-theoretic lengths have been introduced \cite{hall99,hall01}
  \begin{eqnarray}
  \label{eq:shannl}
      l_{Shan} = \exp\left(S[\rho]\right), \quad \text{Shannon length},
  \end{eqnarray}
\begin{eqnarray}
  \label{eq:fisherl}
    l_{Fish}=\frac{1}{\sqrt{F[\rho]}}\, , \quad \quad \text{Fisher length}.
\end{eqnarray}
It is straightforward to observe that these two lengths, as well as the standard deviation or Heisenberg length $l_{Heis} \equiv \Delta x$, have the same units of $X$. Moreover, all of them scale linearly with $X$, are invariant under translations of X, and vanish in the limit as $\rho(x)$ approaches a delta function.\\

Let us note that the quantities ($V[\rho]$, $S[\rho]$, $D[\rho]$, $F[\rho]$), and its related measures
($l_{Heis}$, $l_{Shan}$, $l_{Fish}$), are complementary since each of them grasps a different single facet of the
probability density $\rho(x)$. So, the variance measures the concentration of the density around the centroid while the disequilibrium and the Shannon entropy are measures of the extent to which the density is in fact concentrated, and the Fisher information is a quantitative estimation of the 
oscillatory character of the density since it measures the pointwise concentration of the probability over its support interval $\Lambda$. The disequilibrium and the Shannon and Fisher quantities are considered to be quantitative measures of the departure from uniformity, total spreading and gradient content  of the density, respectively.\\

Moreover, while the Heisenberg length is a measure of separation of the region(s) of the probability concentration with respect to a particular point of the density (namely, the mean value), both Shannon and Fisher lengths are measures of the extent to which the probability density is in fact concentrated. The Fisher length has two further distinctive features. First, it can be considered as a measure of the length scale over which $\rho (x)$ varies rapidly. Second, it depends on the derivative of the density, so that it vanishes for discontinuous distributions and it is very sensitive to fluctuations of the density. For completeness, let us also collect here that these three lengths satisfy the two following inequality-type relations: 
 \begin{eqnarray}
 \label{eq:ineq1}
 l_{Fish} \le  l_{Heis},\\
  \label{eq:ineq2}
\left(2\pi e \right)^{1/2}l_{Fish} \,  \le \, l_{Shann} \, \le \, \left(2\pi e \right)^{1/2} l_{Heis},
\end{eqnarray}
where the equalities are reached for the Gaussian distribution.\\

Recently, composite density-dependent information-theoretic quantities have been introduced: the complexity measures of Cr\'amer-Rao,
Fisher-Shannon and L\'opez-Ruiz-Mancini-Calbet (LMC) types. They are given by the product of two of the previous single spreading measures
as
  \begin{eqnarray}
  \label{cramerrao}
      C_{CR}[\rho]=F[\rho] \times V[\rho],
  \end{eqnarray}
  
  \begin{eqnarray}
     \label{fishershannon}
      C_{FS}[\rho]=F[\rho] \times \frac{1}{2 \pi e} e^{2 S[\rho]}=\frac{1}{2 \pi e} F[\rho] \times l_{Shan}^2,
  \end{eqnarray}
  
  \begin{eqnarray}
     \label{lmc}
     C_{LMC}[\rho] = D[\rho] \times e^{S[\rho]}=\langle \rho \rangle \times l_{Shan},
  \end{eqnarray}
for the Cr\'amer-Rao \cite{dembo,dehesa_1,antolin_ijqc09}, Fisher-Shannon \cite{romera_1,angulo_pla08} and LMC complexities \cite{catalan_pre02}, respectively. Each of them grasps the combined balance of two different facets
of the probability density. The Cr\'amer-Rao complexity quantifies the gradient content of $\rho(x)$ jointly with the probability
spreading around the centroid. The Fisher-Shannon complexity measures the gradient content of $\rho(x)$ together with its total extent 
in the support interval. The LMC complexity measures the combined balance of the average height of $\rho(x)$ (as given by the disequilibrium $D[\rho]$), and its total extent (as given by the Shannon entropic power or Shannon length
$l_{Shan}=e^{S[\rho]}$). Moreover, it may be observed that these three complexity measures are (a) dimensionless, (b) bounded from
below by unity (when $\rho$ is a continuous density in $\mathbb{R}$ in the Cr\'amer-Rao and Fisher-Shannon cases, and for any $\rho$ in the LMC case)\cite{guerrero}, and (c) minimum for the two extreme (or least complex) distributions which correspond to perfect order (i.e. the extremely
localized Dirac delta distribution) and maximum disorder (associated to a highly flat distribution). Finally, they fulfill invariance properties
under replication, translation and scaling transformation \cite{yamano_jmp04,yamano_pa04}.

\section{Entropy measures of a d-dimensional blackbody}

In this section we determine the main quantifiers of the frequency spreading of the spectral density of a multidimensional blackbody, $\rho_{T}^{(d)}(\nu), d > 1$, in terms of the space dimensionality d and temperature T; namely, variance, disequilibrium, Shannon entropy and Fisher information. According to the previous section, we first normalize to unity the blackbody density by finding that the associated normalization constant (which corresponds to the total energy) is
\begin{equation}
a_{d} = \frac{2 (d-1) \left(\frac{\sqrt{\pi }}{c}\right)^d \zeta (d+1) \Gamma (d+1) (k_{B} T)^{d+1}}{h^d \,\Gamma \left(\frac{d}{2}\right)},
\end{equation}
so that the normalized d-dimensional Planck radiation formula is given by 
\begin{equation}
\rho(\nu) \equiv \frac{\rho_{T}^{(d)}(\nu)}{a_{d}} = \frac{1}{\Gamma(d+1)\zeta(d+1) }\left(\frac{h}{k_{B}T}\right)^{d+1}\frac{\nu^{d}}{e^{\frac{h \nu }{k_{B} T}}-1}, 
\end{equation}
where $\zeta (x)$ denotes the zeta function of Riemann \cite{olver}.

\subsection{ Variance}
Since the expressions for the averages of $\nu$ and $\nu^{2}$ are given by
\begin{eqnarray}
\label{eq:nuavenuavequad}
\label{eq:nuavequad}
\langle \nu \rangle &=& (d+1)\frac{ \zeta (d+2)}{ \zeta (d+1)}\frac{k_{B} T}{h} \\
\langle \nu^{2} \rangle &=& (d+1) (d+2)  \frac{\zeta (d+3)}{ \zeta (d+1)}\left(\frac{k_{B} T}{h}\right)^{2},
\end{eqnarray}
one has that the variance of the d-dimensional blackbody radiation density $\rho(\nu)$ is 
\begin{eqnarray}
\label{eq:varian}
V(d,T) \equiv V^{(d)} [\rho] = \langle \nu^{2}\rangle-\langle \nu\rangle^{2} = C_{1}(d)\left(\frac{k_{B} T}{h}\right)^{2},
\end{eqnarray}
with
\begin{eqnarray}
C_{1}(d) = \frac{(d+1) \left((d+2) \zeta (d+1) \zeta (d+3)-(d+1) \zeta (d+2)^2\right)}{ \zeta (d+1)^2}.
\end{eqnarray}
 In particular, for $d = 3$ one finds that the variance of the 3-dimensional blackbody radiation density grows quadratically with temperature as given by
\begin{eqnarray}
V(3,T) &=&\frac{40 \left(\pi ^{10}-68040\, \zeta (5)^2\right)}{21 \pi ^8 }\left(\frac{k_{B}T}{h}\right)^{2}\nonumber \\
&\simeq &  4.11326  \left(\frac{k_{B}T}{h}\right)^{2}
\end{eqnarray}

\subsection{Disequilibrium}
The disequilibrium of the d-dimensional blackbody radiation density $\rho(\nu)$ is, according to Eq. \eqref{eq:diseq},
\begin{eqnarray}
\label{eq:diseq2}
D(d,T) \equiv D^{(d)}[\rho] =\langle \rho \rangle =\int_{0}^{\infty}\rho(\nu)^{2}\, d\nu = C_{2}(d) \frac{h}{k_{B} T}, 
\end{eqnarray}
where 
\begin{eqnarray}
 C_{2}(d) = \frac{\Gamma(1+2d)}{\Gamma(1+d)^{2}\zeta(1+d)^{2}}(\zeta (2 d)-\zeta (2 d+1)).
 \end{eqnarray}
Then, the disequilibrium for 3-dimensional blackbody radiation will be 
\begin{eqnarray}
D(3,T) &=& \frac{1200 \left(\pi ^6-945\, \zeta (7)\right)}{7 \pi ^8}\frac{h}{k_{B} T}\nonumber \\
&\simeq &  0.153553 \,\,\frac{h}{k_{B} T}
\end{eqnarray}

\subsection{Shannon entropy}
The Shannon entropy of the d-dimensional blackbody radiation density $\rho(\nu)$ is, according to Eq. \eqref{eq:shann},
\begin{eqnarray}
\label{eq:shann2}
S(d,T) \equiv S^{(d)}[\rho] = -\int_{0}^{\infty}\rho(\nu)\log \rho(\nu)\, d\nu = -\log \frac{h}{k_{B}T} + C_{3}(d)
\end{eqnarray}
where
\begin{eqnarray*}
C_{3}(d) &=& \log [\Gamma(1+d)\zeta(1+d)]-\frac{I(d)}{\Gamma(1+d)\zeta(1+d)}\nonumber 
\end{eqnarray*}
where I(d) denotes the function
\begin{eqnarray*}
I(d) &=& \int_{0}^{\infty} \frac{x^{d}}{e^{x}-1}\log \left(\frac{x^{d}}{e^{x}-1}\right)\, dx \\ 
&=& d\,\Gamma(1+d)(\psi^{(0)}(1+d)\zeta(1+d)+\zeta'(1+d))\nonumber \\
& & -\Gamma(2+d) \zeta(2+d)+\Gamma(1+d)\left(\frac{d+1}{2}\zeta(d+2)-\sum_{k=1}^{d-1}\zeta(d+1-k)\zeta(1+k) \right), \nonumber 
\end{eqnarray*}
Then, the Shannon entropy of the 3-dimensional blackbody radiation is
\begin{eqnarray}
S(3,T) \simeq  2.03655 -\log \left(\frac{ h}{k_{B} T}\right)
\end{eqnarray}


\subsection{ Fisher information}
The Fisher information of the d-dimensional blackbody radiation density $\rho(\nu)$ is, according to Eq. \eqref{eq:fisher},
\begin{eqnarray}
\label{eq:fisher2}
F(d,T) \equiv F^{(d)}[\rho] = \int_{0}^{\infty} \frac{[\rho'(\nu)]^{2}}{\rho(\nu)} \, d\nu = C_{4}(d) \left(\frac{h}{k_{B}T}\right)^{2}
\end{eqnarray}
with
\begin{eqnarray}
C_{4}(d) = \frac{J(d)}{\Gamma(d+1)\zeta(1+d)}
\end{eqnarray}
where J(d) is the function defined by the integral representation
\begin{eqnarray}
J(d) = \int_{0}^{\infty} x^{d-2}\frac{[d(1-\exp(x))+x\exp(x)]^{2}}{(\exp(x)-1)^{3}} \nonumber 
\end{eqnarray}

Then, the Fisher information of the 3-dimensional blackbody radiation is
\begin{eqnarray}
F(3,T) &=& \left(\frac{h}{k_{B}T}\right)^{2}\frac{15}{\pi^{4}}J(3)\nonumber \\
&\simeq & \left(\frac{h}{k_{B}T}\right)^{2}\frac{15}{\pi^{4}} 3.60617\nonumber \\
&\simeq &  0.555313 \left(\frac{h}{k_{B}T}\right)^{2}.
\end{eqnarray}

\section{Characteristic frequencies of a d-dimensional blackbody}
The most characteristic frequency of the Planck radiation spectrum in a d-dimensional universe is $\nu_{max}(d,T)$, the frequency where the density reaches the maximum. It is well known that this frequency, according to the generalised Wien displacement law \cite{cardoso, ramos1}, is proportional to temperature as 
\begin{eqnarray}
\label{eq:wien}
\nu_{max}(d,T) &=& C_{0} (d)\, \frac{k_{B} T}{h},
\end{eqnarray}
with the constant
\[C_{0}(d) = d + W(-d e^{-d}),
\]
where the Lambert function $W(z)$ is defined \cite{olver} as $W(z) + e^{W(z)} = z$. Then, taking into account the properties of this function, the constant $C_{0}(d)$ can be shown to increase as the dimensionality is increasing. Let us also collect here that for $d = 3$, the constant $C_{0}(3)  \simeq  2.82144$.\\

Here, following the indications given in section II, we introduce for the first time to the best of our knowledge three new characteristic frequencies of the d-dimensional Planck spectrum based on the spreading measures of the Planck density discussed in the previous section; namely, the variance, the Shannon entropy and the Fisher information. They are referred as Heisenberg, Shannon and Fisher frequencies, respectively, for obvious reasons.\\

\begin{center}
\begin{figure}[h]
\includegraphics[scale=0.75]{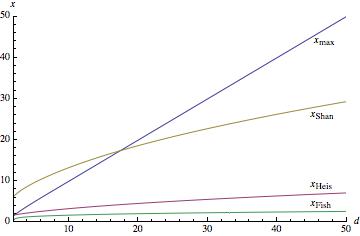}
\caption{Wien-like laws of the characteristic frequencies $x_{i}(d)$ for the d-dimensional blackbody radiation }
\label{fig:figura1}
\end{figure}
\end{center}

The Heisenberg, Shannon and Fisher frequencies of the d-dimensional Planck frequency density are given, according to the corresponding notions defined in section II and Eqs. \eqref{eq:varian}, \eqref{eq:shann2} and \eqref{eq:fisher2}, by 
\begin{eqnarray}
\label{eq:standevi}
\nu_{Heis}(d,T) &=& \sqrt{V(d,T)} = \sqrt{C_{1}(d)}\,\,\frac{k_{B}T}{h},
\end{eqnarray}
\begin{eqnarray}
\label{eq:shannlength}
\nu_{Shan}(d,T) &=& \exp\left(S(d,T)\right) = \exp\left(C_{3}(d)\right) \frac{k_{B}T}{h},
\end{eqnarray}
\begin{eqnarray}
\label{eq:fisherlength}
\nu_{Fish}(d,T) &=& \frac{1}{\sqrt{F(d,T)}}
					 = \frac{1}{\sqrt{C_{4}(d)}}\,\frac{k_{B}T}{h},
\end{eqnarray}
respectively. Note that these three frequencies follow a displacement law similar to the generalised Wien law \eqref{eq:wien} fulfilled by the frequency $\nu_{max}(d,T)$ where the spectrum is maximum. Indeed they do depend linearly on the temperature, and the proportionality constant only depends on the universe dimensionality. The relative comparison among these four Wien-like laws is done in Figure \ref{fig:figura1}, where the dimensionless quantity
\begin{eqnarray}
x_{i} &=& \frac{h}{k_{B}T}\nu_{i},   \quad \text{for i = max, Heis, Shan, Fish},
\end{eqnarray}
has been plotted as a function of the dimensionality d.

We observe the following relative behaviour among the four characteristics frequencies of the d-dimensional blackbody spectrum:
\begin{eqnarray}
 x_{Fish}<x_{max}<x_{Heis}<x_{Shan}&, &  d=2 \\
x_{Fish}<x_{Heis}<x_{max}<x_{Shan}&, & d\leq 17\\
 x_{Fish}<x_{Heis}<x_{Shan}<x_{max} &, & d\geq 18 
\end{eqnarray}
It is worth noting that this behaviour is universally valid in the sense that it holds for any value of the absolute temperature of the system. Let us also mention that for $d\geq 13$, $x_{max}$ grows linearly with $d$. Moreover, it is trivial to show that $\left(2\pi e \right)^{1/2}x_{Fish} \, < \, x_{Shan} < \left(2\pi e \right)^{1/2} x_{max}$ so that the general inequality \eqref{eq:ineq2} is fulfilled, what is a further checking of our results.\\
 
On the other hand, for $d=3$ one has that the variance-based Heisenberg frequency and the entropy-based Shannon and Fisher frequencies of the 3-dimensional blackbody radiation are given by
\begin{eqnarray}
\label{eq:stdevd3}
\nu_{Heis}(3,T) &=& \frac{180}{\pi^{4}}\left( \frac{\pi ^{10}}{17010}-4 \,\zeta (5)^2   \right)^{1/2} \frac{k_{B}T}{h}  \simeq 2.02812 \,\frac{k_{B}T}{h} \quad ,
\end{eqnarray}

\begin{eqnarray}
\nu_{Shan}(3,T) &=& \exp\left(C_{3}(3)\right) \frac{k_{B}T}{h} \simeq 7.66411 \frac{k_{B} T}{h},
\end{eqnarray}

\begin{eqnarray}
\nu_{Fish}(3,T) &=&   \frac{1}{\sqrt{C_{4}(3)}}\,\frac{k_{B}T}{h} \simeq 1.34193\, \frac{k_{B}T}{h},
\end{eqnarray}	respectively. Then, in particular, for the temperature $T=2.725 \text{K}$ one finds that $\nu_{max}(cmb) = 1.60201\cdot 10^{11}$Hz and the following values  
\begin{equation}
\label{eq:stdevd3T2725}
\nu_{Fish}(cmb)  \simeq 6.57748\cdot 10^{10} \, \text{Hz}, \quad \nu_{Heis}(cmb) \simeq 1.15156\cdot 10^{11} \, \text{Hz},\quad \nu_{Shan}(cmb) \simeq 4.35167 \cdot 10^{11}\,\text{Hz}.
\end{equation}

%

for the Heisenberg, Shannon and Fisher frequencies of the 3-dimensional cosmic microwave background (cmb, in short) radiation. It is interesting to realise that the Fisher frequency has the lowest value among all the four characteristic spectral frequencies, partially reflecting the high degree of smoothness of the cosmic microwave spectrum. Since tiny fluctuations of $\rho(\nu)$ can cause huge changes in the Fisher frequency but hardly alter the other three characteristic frequencies of the spectrum, this quantity might be paradoxically the most important frequency to detect the small deviations of the Planck radiation formula which have been observed \cite{mather} in the cosmic microwave radiation. These deviations have been argued to be due to the interaction between photons and other microscopic particles and/or to the nonextensive statistics environment which is presumably associated to the long-range interactions; much work along these lines are being done in the last few years \cite{zeng,arplastino, martinez,valluri}. Then, we are led to conjecture that the Fisher frequency is an appropriate quantifier of the anisotropies of the cosmic microwave radiation that should be calculated in the framework of the nonlinear models \cite{zeng} and non-extensive theories \cite{tsallis} of the cosmic background radiation recently proposed, and it might be experimentally determined.

%
%

\section{Complexity measures of a d-dimensional blackbody}
In this section we calculate the three main intrinsic complexity measures of the  energy density of the d-dimensional blackbody radiation; namely, the Crámer-Rao, Fisher-Shannon and LMC quantities. The Crámer-Rao measure quantifies the frequency-gradient content of the spectrum distribution jointly with the frequency concentration around its mean value, while the Fisher-Shannon complexity measures the frequency-gradient content together with its spreading. On the other hand, the LMC complexity measures the combined balance of the average height of the spectrum and its effective extent.

Taking into account the corresponding notions defined in section II and Eqs. \eqref{eq:varian}, \eqref{eq:diseq2}, \eqref{eq:shann2} and \eqref{eq:fisher2}, the Crámer-Rao, Fisher-Shannon and LMC complexity measures of the d-dimensional Planck frequency density are given by
\begin{eqnarray}
 \label{eq:crrao}
	C_{CR}(V,T) &=& F(d,T) \cdot V(d,T) = C_{4}(d) \,C_{1}(d), \\
	C_{LMC}(d,T) &=& D(d,T) \, \,  e^{S(d,T)}	= C_{2}(d) \, e^{C_{3}(d)},\\
 C_{FS}(d,T) &=& F(d,T)\cdot \frac{1}{2\pi \, e}e^{2\,S(d,T)} = C_{4}(d) \frac{1}{2\pi \, e} e^{2C_{3}(d)},
\end{eqnarray}
respectively. Moreover, taking into account the dimensionless constants $C_{i}$, with i = 1, 2, 3 \, and\, 4, given in the previous section, one finds the following values
\begin{eqnarray}
C_{CR}(d,T) =\frac{(d+1) \left((d+2) \zeta (d+1) \zeta (d+3)-(d+1) \zeta (d+2)^2\right)}{\zeta (d+1)^3 \Gamma (d+1)}\times J(d),   \\
C_{LMC}(d,T) =  \frac{\Gamma(1+2d) (\zeta (2 d)-\zeta (2 d+1))}{\Gamma(1+d)\zeta(1+d)}\, \exp\left(-\frac{I(d)}{\Gamma(1+d)\zeta(1+d)}\right),   \\
C_{FS}(d,T) =  \frac{1}{2\pi e}\Gamma(d+1)\zeta(1+d) \, J(d) \exp\left(-\frac{2\,I(d)}{\Gamma(1+d)\zeta(1+d)}\right)
\end{eqnarray}
for the Crámer-Rao, Fisher-Shannon and LMC complexity measures of the d-dimensional blackbody radiation, respectively.

\begin{center}
\begin{figure}[h]
\includegraphics[scale=0.75]{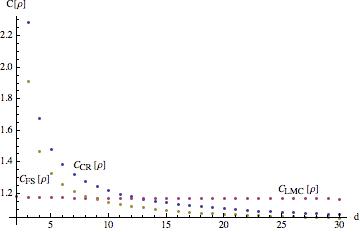}
\caption{The Crámer-Rao, Fisher-Shannon and LMC complexity measures of the d-dimensional blackbody radiation in terms of d.}
\label{fig:figura2}
\end{figure}
\end{center}

Interestingly, we notice that all three complexities of the d-dimensional blackbody radiation depend neither on temperature nor on the Planck and Boltzmann constants, but they only depend on dimensionality. The former result indicates that these three measures of complexity have an universal character for any d-dimensional blackbody; at least in principle, that is without taking into account possible physical processes which might slightly modify the blackbody spectrum, such as the photon scattering by high-energy electrons, quantum gravity effects and non-extensivity effects of the medium, among others. 
The dimensionality dependence of the complexity measures is shown in Figure \ref{fig:figura2}. 

Therein we observe that the three measures of complexity monotonically decrease with different slopes when the dimensionality is increasing. The LMC complexity is almost constant when $d\ge 2$. The other two measures of complexity clearly decrease, mainly because of the Fisher component included in both quantities. The Fisher-Shannon measure decreases faster than the Crámer-Rao one, basically because the Shannon component included in the former measure is bigger than the variance component of the latter complexity. This is the same phenomenon which explains that the Shannon frequency is always bigger than the Heisenberg frequency, as discussed in the previous section. On the other hand, let us highlight that the Fisher-Shannon and Crámer-Rao measures of complexity can be used as quantifiers of the space dimensionality as grasped by the blackbody spectrum. Moreover, they drastically vary even with small dimensional oscillations of the spectrum because of the huge variations of their Fisher ingredient, as already mentioned. From this point of view, these two measures of complexity could be eventually used to detect dimensional anisotropies of the spectrum. Let us finally note that for $d = 3$ one finds the values 1.17685, 1.90979, 2.28415 for the LMC, Fisher-Shannon and Crámer-Rao measures of complexity of the 3-dimensional blackbody radiation.

\section{Conclusions}
In this paper we have investigated the entropy and complexity quantities of the d-dimensional blackbody radiation for standard ($d = 3$) and non-standard dimensionalities. We have calculated the main entropy and complexity measures of the corresponding spectral energy density in terms of dimensionality
d and temperature T. First, we have determined the variance, the disequilibrium, the Shannon entropy and the Fisher information of the d-dimensional Planck density in an explicit way. Then, besides the frequency $\nu_{max}$ at which the density is maximum, we have used these frequency-spreading measures to introduce three further characteristic frequencies of the spectrum, which have been referred as Heisenberg, Shannon and Fisher frequencies for obvious reasons. We have found that these new frequencies have a dependence on the temperature similar to the one given by the well-known Wien's law followed by $\nu_{max}$. The values of these charateristic frequencies for the cosmic microwave background radiation have been given and physically discussed, suggesting the potential interest of the Fisher frequency to grasp the CMB anisotropy.\\

Second, we have shown that the three main measures of complexity (i.e., Crámer-Rao, Fisher-Shannon and LMC) do not depend on the temperature, but only on the universe dimensionality. The corresponding values for the cosmic microwave radiation turns out to be dimensionless constants, noting that the Crámer-Rao complexity is bigger than the Fisher-Shannon and LMC quantities mainly because of the high smoothness and extent of the corresponding 3-dimensional Planck density. On the other hand, it is worth remarking the potential quality of these complexities (mainly the Crámer-Rao and Fisher-Shannon) as identifiers of CMB anisotropies because of the sensitivity of the Fisher-information component to the fluctuations of the associated non-smoothed Planck density due to (a) different physical processes such as the photon scattering by high-energy electrons (Sunyaev-Zel'dovich effect) or gravitational shift of photon energy caused by varying gravitational fields (Sachs-Wolfe effect), and (b) cavity finite-size corrections \cite{garcia, ramos2}.\\

A possible extension of this work is the inclusion of quantum gravity effects, which will certainly modify the black body spectrum and might open new windows to know deeper the quantum gravitational features of the early universe through study of CMB spectrum. This could be done by taking into account the quantum gravity effects encoded in modified dispersion relations \cite{nozari}.\\

Finally, let us point out that this entropy and complexity analysis should be extended as well to the nonlinear blackbody radiation laws \cite{zeng,tsallis}, which presumably takes into account the small deviations from the Planck radiation formula that have been detected \cite{mather} in the cosmic microwave radiation. The origin of these deviations has been argued to be due both to the interaction between photons and other microcosmic particles \cite{zeng} and the nonextensive statistics environment \cite{tsallis} which is associated with the long-range interactions. The last approach has received much attention (see e.g., \cite{arplastino,martinez,valluri} and references therein).

\acknowledgments This work was partially supported by the Projects
FQM-2445 and FQM-207 of the Junta de Andalucia and the grant
FIS2011-24540 of the Ministerio de Innovaci\'on y Ciencia (Spain).

\end{document}